\documentstyle[pre,aps]{revtex}
\begin{document}
\draft
\title
{A discretized integral hydrodynamics}
\author{
V\'{\i}ctor Romero-Roch\'{\i}n\footnote{Permanent address: Instituto de 
F\'{\i}sica, UNAM.
Apdo. Postal 20-364, 01000 M\'exico, D.F. Mexico.} and J. Miguel Rub\'{\i}}
\address{
Departament de F\'{\i}sica Fonamental.
Universitat de Barcelona. \\
Av. Diagonal 640, E-08028 Barcelona, Spain.}
\maketitle

\begin{abstract}
Using an interpolant form for the gradient of a function of position,
we write an integral version of the conservation equations for a fluid.
In the appropriate limit, these become the usual conservation laws of
mass, momentum and energy. We also discuss the special cases of the
Navier-Stokes equations for viscous flow and the Fourier law for thermal 
conduction in
the presence of hydrodynamic fluctuations.
By means of a discretization procedure, we show how these equations can give
rise to the so-called ``particle dynamics" of Smoothed Particle 
Hydrodynamics and Dissipative Particle Dynamics.
\end{abstract}

\section{Introduction}

The inherent difficulties of the equations for fluid hydrodynamics
has given rise to a variety of schemes that numerically simulate the
behavior of a fluid at hydrodynamic scales \cite{simula}. Of renewed
interest, there have appeared the simulation schemes based on the idea of 
substituting the fluid by ``particles" that move under the influence
of forces that, in a coarse-grain limit, reduce to some approximate
form of the hydrodynamic Navier-Stokes equations; these schemes are
called Smoothed Particle 
Hydrodynamics (SPH) \cite{SPH1,SPH2} and Dissipative Particle Dynamics (DPD)
\cite{DPD}. Although
their numerical implementation is somewhat different, the latter including
a random force, we shall argue that their origin is essentially the same.
The use of these particle-like simulations has been reported of
being successful in different applications of fluid dynamics 
\cite{ESPH,liver,DPD2,EDPD,Spanish,Ernst}.
An attractive feature of these simulations is that 
one can use the enormous experience gained from the techniques
of simulations of standars molecular dynamics; in particular, in
dealing with rheological fluids.

In this article, we present an integral representation of the
hydrodynamic conservation laws based on the concept of the
interpolant of a function \cite{SPH1,SPH2}; the interpolant is
an integral representation of a function in terms of a weighting
function. This function, in the appropriate limit, becomes a delta function 
and the interpolant yields an identity. For calculational 
purposes one
does not take such a limit and, therefore, one ends with an
approximating form of the corresponding function.  The SPH simulation
is based on a certain form of interpolant of the hydrodynamic
{\it functions}, such as density, velocity, etc. Here, we base
our scheme not on the interpolants of the functions, but rather on
the interpolants of the {\it gradient} of the functions. This minor 
change proves to be very useful in writing integral equations that 
reduce, exactly,
to the conservation laws in the limit. Moreover, we shall see
that the conservation of mass, momentum and energy is exactly preserved
at the level of the integral forms. At an approximation level,
we shall show that the integral forms of the conserving currents 
are correct up to second order terms in the gradients of the
corresponding fields. 
Since the integral expressions can be written down
following the forms of the true laws, the phenomenological 
variables, such as viscosities and thermal conductivity, can be
naturally included. In the same fashion, the extension to fluctuating
hydrodynamics can be readily performed.

The resulting integral conservation laws may then be used as
an alternative to the exact differential laws. Further, they
can be utilized as the starting point for approximate solutions.
In this context, we show that by an appropriate discretization of
the integrals one can render the equations to look as the equations,
not for the hydrodynamic fields evaluated at space-time points, but
rather, for quantities pertaining to {\it particles}. For instance
the field velocity ${\bf v}({\bf r},t)$ becomes the velocity
${\bf v}_i(t)$ of a particle at position ${\bf r}_i(t)$; a law
of motion for the latter must then be supplied. Within the present
theory one readily finds the equations of motion for the particles
that are fully consistent with the hydrodynamic equations. We shall
show how the SPH and DPD equations may be found. As we shall see, in
general, one can have additional terms arising from the convective
non-linear terms of the hydrodynamic time derivative.

We organize the article as follows. In the next section we introduce
the interpolant of the gradient of a function. With such an object,
we write down the conservation laws, both in general and for the
cases of Navier-Stokes and Fourier law. The extension to fluctuating
hydrodynamics is also shown. In section III we present a discretization
procedure that yields a particle-like simulation algorithm. We discuss
the relationship with the SPH and DPD theories. We conclude and make
additional remarks in section IV.

\section{Hydrodynamics in integral form}

\subsection{An interpolant of the gradient}

The idea of formulating an integral version of the laws
of hydrodynamics is based on a limiting representation of the
gradient of a function of position ${\bf r}$. This representation
we call it an {\it interpolant}, following Lucy \cite{SPH1} and
Monaghan \cite{SPH2} SPH formulation. As we mentioned in the
Introduction, we use the interpolant of the gradient of a 
function rather than that of the function itself. 

First, we show that the following identity is correct,
\begin{equation}
\nabla A({\bf r}) = - {\rm lim}_{r_0 \to 0} \int d{\bf r}^{\prime} \>
A({\bf r}^{\prime}) \> {({\bf r} - {\bf r}^{\prime}) \over 
|{\bf r} - {\bf r}^{\prime}|} W(|{\bf r} - {\bf r}^{\prime}|;r_0) ,
\label{int1}
\end{equation}
where the tensorial character of $A$ is left unspecified, and 
$W(|{\bf r} - {\bf r}^{\prime}|;r_0)$ is a distribution or
weighting function sharply peaked with width $r_0$. We demand 
that all its moments
exist although the function itself may not be integrable. We assume the
following:
\begin{equation}
\int d{\bf r} \> |{\bf r}|^{n} \> W(|{\bf r}|;r_0) = r_0^{n - 1} {\cal M}_n
\label{mom}
\end{equation}
with $n \ge 1$. For $n = 1$ we require ${\cal M}_1 = 3$ but for $n > 1$
we leave ${\cal M}_n$ unspecified. 
These requirements are easily satisfied by 
noticing that the integrand of (\ref{int1}) can be related to the 
gradient of a distribution that tends to a delta function, namely
\begin{equation}
{{\bf r} \over |{\bf r}|} W(|{\bf r}|;r_0) = - \nabla f(|{\bf r}|;r_0)
\label{rela}
\end{equation}
where $f(r;r_0)$ is such, that in the limit $r_0 \to 0$ becomes
\begin{equation}
{\rm lim}_{r_0 \to 0} f(r;r_0) = {1 \over 4 \pi r^2} \delta(r) .\label{delta}
\end{equation}
Clearly, any distribution that tends to $\delta(x)$ can be used. It is
important to stress that $W$ is a function of the magnitude $|{\bf r}|$
and not of the corresponding vector. This dependence is important
for the use of approximations and, as we shall see later on, 
for the setting up of the conservation laws.

 For the validity of
(\ref{int1}) we further require that $A({\bf r})$ is analytic everywhere inside
the domain of the integral. This is not a stringent requirement since we
are interested in hydrodynamic fields. Calling $I(A({\bf r}))$ the integral
in equation (\ref{int1}), we first make a change of variable, 
${\bf r}^\prime \to {\bf r} + {\bf r}^\prime$, and then, we perform a
Taylor expansion of $A({\bf r} + {\bf r}^{\prime})$ around ${\bf r}$. We
obtain
\begin{eqnarray}
I(A({\bf r})) &=& \partial_\alpha A({\bf r}) \int d{\bf r}^{\prime} \>
{r_\alpha^\prime r_\beta^\prime \over 
|{\bf r}^{\prime}|} W(|{\bf r}^{\prime}|;r_0) + \nonumber \\
&& {1 \over 2} \partial_\alpha \partial_\gamma \partial_\eta A({\bf r})
\int d{\bf r}^{\prime} \> {r_\alpha^\prime r_\beta^\prime r_\gamma^\prime
r_\eta^\prime \over |{\bf r}^{\prime}|} W(|{\bf r}^{\prime}|;r_0) + \dots,
\label{int2}
\end{eqnarray}
where $\partial_\alpha = \partial / \partial r_\alpha$ with $r_\alpha$
the cartesian components of the vector ${\bf r}$. Due to the spherical 
symmetry of $W$ all the odd powers of
${\bf r}^\prime$ (even powers in the derivatives) vanish identically. With
the use of Eq. (\ref{mom}) the first integral in the rhs of (\ref{int2})
gives the Kronecker delta $\delta_{\alpha \beta}$, independently of $r_0$,
while the second integral yields $r_0^2 (\delta_{\alpha \beta} 
\delta_{\gamma \eta} + \delta_{\alpha \gamma} \delta_{\beta \eta}
+ \delta_{\alpha \eta} \delta_{\beta \gamma}) {\cal W}_3 / 15$. Clearly the
higher order terms are proportional to $r_0^{2n}$ times odd 
($2n + 1$)-derivatives of $A$. We write, generically,
\begin{equation}
I(A({\bf r})) = \nabla A({\bf r}) + {\cal O}\left( r_0^2 \nabla^3 A \right) .
\label{int3}
\end{equation}
In the limit $r_0 \to 0$, the interpolant is the gradient of $A$. For
approximation purposes, we note from (\ref{int3}) that the correction to the 
gradient is third order in the derivatives.

\subsection{Conservation laws}

The study of the hydrodynamics of a fluid is based on the 
conservation laws for mass, momentum an energy. \cite{LL-H} 
In the following we assume that all the fields are evaluated at a spatial
point ${\bf r}$ and that all are time dependent as well. The conservation
laws are
\begin{eqnarray}
{\partial \rho \over \partial t} &=& - \nabla \cdot \left( \rho {\bf v} \right),
\label{mass}
\\
{\partial {\bf j} \over \partial t} &=& - \nabla \cdot \left(
{\bf j v} + P \tilde 1 - \tilde \Pi \right)  \label{momen}
\\
{\partial e \over \partial t} &=& - \nabla \cdot \left( e {\bf v}
+ (P \tilde 1 - \tilde \Pi) \cdot {\bf v} + {\bf J} \right) 
\label{ener}
\end{eqnarray}
where $\rho$ is the mass density; ${\bf v}$ is the velocity of the
fluid; ${\bf j}$ is the momentum density, ${\bf j} = \rho {\bf v}$;
$P$ is the hydrostatic pressure; $\tilde \Pi$ is the viscous
stress tensor; $e = \rho {\bf v}^2 /2 + u$ is the total energy 
density with $u$ the internal energy density of the fluid; and ${\bf J}$ 
is the heat current. Integration over the whole volume of the fluid shows that
the total mass, momentum and energy of the fluid are conserved.

With the use of the interpolant given by Eq. (\ref{int1}), we can write
down analogous conservation expressions (with implicit time-dependence):
\begin{eqnarray}
{\partial \rho({\bf r}) \over \partial t} &=&  \int d{\bf r}^\prime
W(|{\bf r} - {\bf r}^{\prime}|;r_0){({\bf r} - {\bf r}^{\prime})_\alpha \over 
|{\bf r} - {\bf r}^{\prime}|}  
\left[ \rho({\bf r}^\prime) v_\alpha({\bf r}) +  
\rho({\bf r}) v_\alpha({\bf r}^\prime)\right]
\label{Imass} \\
\> \nonumber \\
{\partial j_\beta({\bf r}) \over \partial t} &=&  \int d{\bf r}^\prime
W(|{\bf r} - {\bf r}^{\prime}|;r_0){({\bf r} - {\bf r}^{\prime})_\alpha \over 
|{\bf r} - {\bf r}^{\prime}|} 
\{ \left[ j_\beta({\bf r}^\prime) v_\alpha({\bf r}) 
+ j_\beta({\bf r}) v_\alpha({\bf r}^\prime)\right] \nonumber \\
&&+ \left[ P({\bf r};{\bf r}^\prime) +  P({\bf r}^\prime;{\bf r}) \right] 
\delta_{\alpha \beta} 
 - \left[ \Pi_{\alpha \beta}({\bf r};{\bf r}^\prime) +  
\Pi_{\alpha \beta}({\bf r}^\prime;{\bf r}) \right] \}
\label{Imom}
\\
\> \nonumber \\
{\partial e({\bf r}) \over \partial t} &=&  \int d{\bf r}^\prime
W(|{\bf r} - {\bf r}^{\prime}|;r_0)
{({\bf r} - {\bf r}^{\prime})_\alpha \over 
|{\bf r} - {\bf r}^{\prime}|} \{
\left[ e({\bf r}^\prime)  v_\alpha({\bf r}) +  
e({\bf r}) v_\alpha({\bf r}^\prime)\right] 
\nonumber \\
&& + 
\left[ P({\bf r};{\bf r}^\prime)v_\alpha({\bf r}) +  
P({\bf r}^\prime;{\bf r}) v_\alpha({\bf r}^\prime) \right] 
- \left[ \Pi_{\alpha \beta}({\bf r};{\bf r}^\prime) v_\beta({\bf r}) +  
\Pi_{\alpha \beta}({\bf r}^\prime;{\bf r}) v_\beta({\bf r}^\prime)
\right] \nonumber \\
&&+ \left[ J_\alpha({\bf r};{\bf r}^\prime) +  
J_\alpha({\bf r}^\prime;{\bf r}) \right] \} .
\label{Iene}
\end{eqnarray}

The above expressions, by the rule of the interpolant, are equal (in the
limit $r_0 \to 0$) to 
(minus) the divergence of the terms in square brackets evaluated at
${\bf r}^\prime = {\bf r}$; in particular, the ``kernels"
$P({\bf r};{\bf r}^\prime)$, $\tilde \Pi({\bf r};{\bf r}^\prime)$
and ${\bf J}({\bf r};{\bf r}^\prime)$ must be chosen such
that the interpolant equal the divergence of the actual pressure, 
viscous stress tensor and heat current when ${\bf r}^\prime = {\bf r}$. 
But before we discuss
how to choose these kernels we point out that all the terms inside
the square brackets are {\it symmetric} with respect to the interchange
of the variables ${\bf r}$ and ${\bf r}^\prime$. Therefore, integration
with respect to ${\bf r}$ makes the rhs of all the equations vanish,
thus yielding the conservation of the extensive variables,
independently of whether the limit $r_0 \to 0$ is taken or not.

\subsection{Constitutive relations}

The conservation laws must be provided with constitutive relations
in order to have a closed set of equations. For discussion purposes
we shall choose the mass density, velocity and temperature $T({\bf r},t)$
as the independent fields. Therefore, for the pressure and the
internal energy we need
to know the equations of state of the fluid in terms of $\rho$ and $T$;
in particular, we assume we know the functional dependence
$P(\rho,T)$. Thus, the kernel for the
pressure may be chosen as
\begin{equation}
P({\bf r};{\bf r}^\prime) = P\left(\rho({\bf r}),T({\bf r}^\prime) \right).
\label{Pres}
\end{equation}
Clearly, the kernel is not symmetric in its variables and hence
we need the symmetrization in the conservation equations (\ref{Imass}) -
(\ref{Iene}) above. Substitution
of this form into, say, Eq. (\ref{Imom}), gives to lowest order, the term
\begin{eqnarray}
\int d{\bf r}^\prime 
\left[ P({\bf r};{\bf r}^\prime) +  P({\bf r}^\prime;{\bf r}) \right]
{({\bf r} - {\bf r}^{\prime}) \over 
|{\bf r} - {\bf r}^{\prime}|} W(|{\bf r} - {\bf r}^{\prime}|;r_0) = \\
 -\left( {\partial P \over \partial \rho} \right) \nabla \rho
- \left( {\partial P \over \partial T} \right) \nabla T ,\label{Ipress}
\end{eqnarray}
where the rhs is evaluated at ${\bf r}^\prime = {\bf r}$. The rhs is
$- \nabla P({\bf r})$.

Regarding the viscous stress tensor $\tilde \Pi$, we use the usual one that 
gives rise to the Navier-Stokes equations, linear in the velocity gradients.
The corresponding kernel may be written as
\begin{eqnarray}
\Pi_{\alpha \beta}({\bf r};{\bf r}^\prime) &=&
 - \eta({\bf r})
\int d{\bf r}^{\prime \prime} W(|{\bf r}^\prime - {\bf r}^{\prime \prime}|;r_0) 
\left[ {({\bf r}^\prime - {\bf r}^{\prime \prime})_\alpha 
 \over |{\bf r}^\prime - {\bf r}^{\prime \prime}|}
v_\beta({\bf r}^{\prime \prime}) \right. \nonumber \\
&& \left. + {({\bf r}^\prime - {\bf r}^{\prime \prime})_\beta
 \over |{\bf r}^\prime - {\bf r}^{\prime \prime}|}
v_\alpha({\bf r}^{\prime \prime}) - {2 \over 3} \delta_{\alpha \beta}
{({\bf r}^\prime - {\bf r}^{\prime \prime})_\nu
 \over |{\bf r}^\prime - {\bf r}^{\prime \prime}|}
v_\nu({\bf r}^{\prime \prime}) \right] \nonumber \\
&&- \zeta({\bf r})
\int d{\bf r}^{\prime \prime} W(|{\bf r}^\prime - {\bf r}^{\prime \prime}|;r_0)
 \left[ \delta_{\alpha \beta}
{({\bf r}^\prime - {\bf r}^{\prime \prime})_\nu
 \over |{\bf r}^\prime - {\bf r}^{\prime \prime}|}
v_\nu({\bf r}^{\prime \prime}) \right] , \label{Istress}
\end{eqnarray}
where the viscosities $\eta({\bf r})$ and $\zeta({\bf r})$ are, either,
given functions of ${\bf r}$, or they depend on ${\bf r}$ through a 
further dependence on
density and temperature. Substitution of this equation into
Eq.(\ref{Imom}), yields to lowest order the familiar viscosity terms of the
Navier-Stokes equations:
\begin{eqnarray}
- \int d{\bf r}^\prime 
\left[ \Pi_{\alpha \beta}({\bf r};{\bf r}^\prime) +  
\Pi_{\alpha \beta}({\bf r}^\prime;{\bf r}) \right] \cdot
{({\bf r} - {\bf r}^{\prime})_\beta \over 
|{\bf r} - {\bf r}^{\prime}|} W(|{\bf r} - {\bf r}^{\prime}|;r_0) =
\nonumber \\
\partial_\beta \left[ \eta({\bf r}) 
\left( \partial_\alpha v_\beta({\bf r}) + \partial_\beta v_\alpha({\bf r})
- {2 \over 3} \delta_{\alpha \beta} \partial_\nu v_\nu({\bf r}) \right)
+ \zeta({\bf r}) 
\delta_{\alpha \beta} \partial_\nu v_\nu({\bf r}) 
\right]. \label{stress}
\end{eqnarray}
For the form of heat current ${\bf J}$ we consider Fourier law
in which the current is linear in the temperature gradient. The
kernel can be written as,
\begin{equation}
{\bf J}({\bf r};{\bf r}^\prime) = 
\kappa({\bf r})
\int d{\bf r}^{\prime \prime} W(|{\bf r}^\prime 
- {\bf r}^{\prime \prime}|;r_0) 
 {({\bf r}^\prime - {\bf r}^{\prime \prime}) 
 \over |{\bf r}^\prime - {\bf r}^{\prime \prime}|}
T({\bf r}^{\prime \prime}) . \label{Iheat}
\end{equation}
Again, substitution into Eq. (\ref{Iene}) yields, to lowest order,
\begin{equation}
\int d{\bf r}^\prime 
\left[{\bf J}_q({\bf r};{\bf r}^\prime) +  
{\bf J}_q({\bf r}^\prime;{\bf r}) \right] \cdot
{({\bf r} - {\bf r}^{\prime}) \over 
|{\bf r} - {\bf r}^{\prime}|} W(|{\bf r} - {\bf r}^{\prime}|;r_0) =
\nabla \cdot \left[ \kappa({\bf r}) \nabla T({\bf r}) \right], \label{heat}
\end{equation} 
where the ${\bf r}$-dependence of $\kappa$ may be given through
its dependence on temperature and density.

Thus, we have shown a consistent way of presenting an integral
form of the equations
of hydrodynamics, such that, in the appropriate limit yield the
true equations; we note that all the phenomenological coefficients
are readily and unambiguously identified. This is an important point 
since, as we
shall show in the next section, the choice of the functional forms
of the stress tensor and the heat current are, by no means, unique;
that is, we can prescribe different functional forms that in the
limit also give similar expressions to the usual hydrodynamic laws.
We shall defer further discussion of this point to section III.

\subsection{Hydrodynamic fluctuations}
We now turn our attention to the formulation of the study of
hydrodynamic fluctuations. Following Landau and Lifshitz \cite{LL-K,vS}
we limit ourselves to small fluctuations around a given flow,
solution to Eqs. (\ref{mass}) - (\ref{ener}), so that a linearization
in the fluctuations is possible. In keeping with our assumption that the
indepedent variables are the mass density, velocity and temperature
of the fluid, we define the fluctuations as linear deviations
from the flow, that is
\begin{equation}
\rho({\bf r},t) = \rho_0({\bf r},t) + \delta \rho({\bf r},t) 
\label{delro}
\end{equation}
and analogous expressions for ${\bf v}({\bf r},t)$ and
$T({\bf r},t)$. The functions $\rho_0({\bf r},t)$, 
${\bf v}_0({\bf r},t)$ and $T_0({\bf r},t)$ constitute a given
flow, solution to the full non-linear integral equations 
(\ref{Imass}) - (\ref{Iene}), with the expressions (\ref{Ipress}) 
and (\ref{Iheat}).
Now, using expressions such as (\ref{delro}) we can linearize
the integral equations in the fluctuations; for instance, the continuity
equation for the fluctuations, {\it cf.} Eq.(\ref{Imass}), becomes
\begin{eqnarray}
{\partial \delta \rho({\bf r}) \over \partial t} &=& - \int d{\bf r}^\prime
W(|{\bf r} - {\bf r}^{\prime}|;r_0) {({\bf r} - {\bf r}^{\prime}) \over 
|{\bf r} - {\bf r}^{\prime}|} \cdot \nonumber \\
&&\times \left[ \rho_0({\bf r}^\prime) \delta {\bf v}({\bf r}) +  
\delta \rho({\bf r}^\prime) {\bf v}_0({\bf r}) +
\rho_0({\bf r}) \delta {\bf v}({\bf r}^\prime) +  
\delta \rho({\bf r}) {\bf v}_0({\bf r}^\prime)\right] 
\label{Idelmass}
\end{eqnarray}
and similar linear equations for the (partial) time derivatives
of momentum density and energy, $\delta {\bf j}$ and $\delta e$. Next, one
identifies the source of the fluctuations as arising from spontaneous
fluctuations of the stress tensor and the heat current. This is implemented
in the usual way, \cite{LL-K} by adding to the {\it linearized} equations
for the fluctuations of the momentum and energy densities, terms proportional 
to the divergence of a random stress tensor and to the divergence of a 
random heat current, respectively. 

To be precise, we add to the equation
for the fluctuation of the momentum density,
\begin{equation}
\int d{\bf r}^\prime
W(|{\bf r} - {\bf r}^{\prime}|;r_0) {({\bf r} - {\bf r}^{\prime}) \over 
|{\bf r} - {\bf r}^{\prime}|} \cdot \tilde \Pi^R({\bf r},{\bf r}^\prime,t)
\label{Rstress}
\end {equation}
where the tensor $\tilde \Pi^R({\bf r},{\bf r}^\prime,t)$ is a gaussian
random stochastic function, symmetric under interchange of ${\bf r}$ and
${\bf r}^\prime$, with zero mean and with its second moment obeying the
usual fluctuation-dissipation relations,
\begin{eqnarray}
\langle \Pi_{\alpha \beta}^R({\bf r}_1,{\bf r}_2,t)
\Pi_{\gamma \nu}^R({\bf r}_3,{\bf r}_4,t^\prime) \rangle = \nonumber \\
 2 k \{ \left[T_0({\bf r}_1) \eta_0({\bf r}_2) +   
T_0({\bf r}_2) \eta_0({\bf r}_1) \right] 
\left( \delta_{\alpha \gamma} \delta_{\beta \nu} +
\delta_{\alpha \nu} \delta_{\beta \gamma} 
 \right) \nonumber \\
  +  k[T_0({\bf r}_1) (\zeta_0({\bf r}_2) - {2 \over 3} \eta_0({\bf r}_2)) + 
T_0({\bf r}_2) (\zeta_0({\bf r}_1) - {2 \over 3} \eta_0({\bf r}_1)) ] 
\delta_{\alpha \beta} \delta_{\gamma \nu} \}
 \nonumber \\
\times \delta(t - t^\prime) 
\left\{ \delta({\bf r}_1 - {\bf r}_3) \delta({\bf r}_2 - {\bf r}_4)
+ \delta({\bf r}_1 - {\bf r}_4) \delta({\bf r}_2 - {\bf r}_3) \right\} .
\label{flustress}
\end{eqnarray}

In the same fashion, we add to the equation for the fluctuation of the 
energy density,
\begin{equation}
\int d{\bf r}^\prime
W(|{\bf r} - {\bf r}^{\prime}|;r_0) {({\bf r} - {\bf r}^{\prime}) \over 
|{\bf r} - {\bf r}^{\prime}|} \cdot {\bf J}^R({\bf r},{\bf r}^\prime,t)
\label{Jstress}
\end {equation}
where ${\bf J}^R({\bf r},{\bf r}^\prime,t)$ is a gaussian
random stochastic function, symmetric under interchange of ${\bf r}$ and
${\bf r}^\prime$, with zero mean and with second moment
\begin{eqnarray}
\langle J_{\alpha}^R({\bf r}_1,{\bf r}_2,t)
J_{\beta}^R({\bf r}_3,{\bf r}_4,t^\prime) \rangle = \nonumber \\ 
2 k \left(T_0^2({\bf r}_1) \kappa_0({\bf r}_2) + 
T_0^2({\bf r}_2) \kappa_0({\bf r}_1) \right)
\delta_{\alpha \beta} \delta(t - t^\prime)\nonumber \\
\times  
\left[ \delta({\bf r}_1 - {\bf r}_3) \delta({\bf r}_2 - {\bf r}_4)
+ \delta({\bf r}_1 - {\bf r}_4) \delta({\bf r}_2 - {\bf r}_3) \right].
\label{fluheat}
\end{eqnarray}
We recall that the quantities $\delta \rho({\bf r},t)$, 
$\delta {\bf j}({\bf r},t)$, etc. depend from those of 
the underlying flow, $\rho_0({\bf r},t)$, ${\bf j}_0({\bf r},t)$, etc.
but not the other way around.
That is, one first solve for the latter and then one finds the fluctuations.
It is understood that the flow is stable; that is, one should always
have $\delta \rho / \rho_0 \ll 1$, etc. In the next section we shall
also comment how one can include the fluctuations within a particle-like
simulation.

\section{A discretized integral hydrodynamics}
The integral formulation presented
in the previous section is simply an approximate
representation of the usual hydrodynamic laws. Their usefulness resides
on whether their solution may be easier to find than those of
the actual equations or on their amenability for approximations.
In this regard, we recall the approximate SPH and DPD schemes where
a particle-like simulation, similar to a molecular dynamics simulation, 
represents the flow of a continuous fluid. In this section we present a
particular discretization of the integral equations of section II
that may be used as the basis for a simulation in terms of ``fluid particles.''
 
\subsection{A particle-like scheme}

	The basic idea is first to divide space in cells of finite 
size $\Delta V$ and, then, to define the field variables for each
cell. We call ${\bf r}_i$ the position vector of the $i$-th cell, and 
the following list summarizes the variables for such a cell:
\begin{equation}
\begin{array}{cccc}
\rho({\bf r},t) \Delta V & \to &   m_i(t) & {\rm mass} \\
{\bf j}({\bf r},t) \Delta V & \to & {\bf p}_i(t) 
& {\rm momentum} \\
e({\bf r},t) \Delta V & \to & \epsilon_i(t) & {\rm energy} \\
{\bf v}({\bf r},t) \Delta V & \to & {\bf v}_i(t) & {\rm velocity} \\
\end{array} 
\end{equation}

Now, the kernels of pressure and viscous stress 
become ``potentials" of force between the $i$-th and $j$-th cells
while the heat density current is now a ``current" of energy between such
cells:
\begin{equation}
\begin{array}{cccc}
P({\bf r};{\bf r}^\prime;t) \Delta V & \to & {\cal P}_{ij}(t) & {\rm pressure 
\> potential} \\
\Pi^{\alpha \beta}({\bf r};{\bf r}^\prime;t) \Delta V & \to & 
\pi_{ij}^{\alpha \beta}(t) 
& {\rm stress \> potential} \\
J_q^\alpha({\bf r};{\bf r}^\prime;t) \Delta V & \to & 
{\cal J}_{ij}^\alpha(t) & {\rm heat \> current} 
\end{array} 
\end{equation}

The integrals are then discretized by summing over cells directly and
not over labels that localize the cell in a Cartesian grid, that is,
\begin{equation}
\int d {\bf r} \to \sum_i^N \Delta V ,
\end{equation}
where we have assumed there are $N$ cells in the total volume. This
discretization implies a careful choice of the discretized version
of the weighting function $W(|{\bf r} - {\bf r}^\prime|)$. That is,
we cannot simply change ${\bf r}$ by ${\bf r}_i$ and ${\bf r}^\prime$
by ${\bf r}_j$ in the functional form of $W$ since the equations
for the moments, Eq. (\ref{mom}), would not be correct. This is due to
the fact that those results make use of the spherical symmetry of $W$.
Instead, we propose the following discretization that give rise to
the correct moments:
\begin{equation}
W(|{\bf r} - {\bf r}^\prime|) \Delta V \to {\cal W}(r_{ij}) \equiv 
4 \pi r_{ij}^2 (\Delta V)^{1/3} \>
W(r_{ij}) .
\end{equation}
where we have defined $r_{ij} = |{\bf r}_i - {\bf r}_j|$. 
This form also takes into account that $W$ is always part of an integrand.
As a particular example, using as a representation of a delta function,
\begin{equation}
\delta(r) = {\rm lim}_{r_0 \to 0}{1 \over r_0} e^{-r/r_0} ,
\end{equation}
yields for $W(r)$,
\begin{equation}
W(r) = {1 \over 4 \pi r_0^4} \left[ 2\left({r_0 \over r}\right)^3
+ \left({r_0 \over r}\right)^2 \right] e^{-r/r_0}
\end{equation}
and, correspondingly for ${\cal W}(r_{ij})$,
\begin{equation}
{\cal W}(r_{ij}) = {(\Delta V)^{1/3} \over r_0^2} 
\left[ 2\left({r_0 \over r_{ij}}\right) + 1 \right] e^{-r_{ij}/r_0} ,
\end{equation}
which shows that, in discretized form, all the moments but the zeroth
are well defined.

With the above reformulation, and defining 
\begin{equation}
\hat e_{ij} = {({\bf r}_i - {\bf r}_j) \over
|{\bf r}_i - {\bf r}_j|} ,
\end{equation}
the conservation equations look 
as follows. Conservation of mass:
\begin{equation}
{\partial m_i \over \partial t} = \sum_j {\cal W}(r_{ij}) \hat e_{ij} 
\cdot \left[m_i {\bf v}_j + m_j {\bf v}_i \right] . \label{dmass}
\end{equation}
Conservation of momentum:
\begin{eqnarray}
{\partial p_i^\alpha \over \partial t} &=& \sum_j {\cal W}(r_{ij}) 
e_{ij}^\beta \left\{ 
\left( p_i^\alpha v_j^\beta + p_j^\alpha v_i^\beta \right) \right.
\nonumber \\
&& \left. + \delta^{\alpha \beta} \left( {\cal P}_{ij} + {\cal P}_{ji} \right)
- \left( \pi_{ij}^{\alpha \beta} + \pi_{ji}^{\alpha \beta} \right) 
\right\} .\label{dmom}
\end{eqnarray}
Conservation of energy:
\begin{eqnarray}
{\partial \epsilon_i \over \partial t} &=& \sum_j {\cal W}(r_{ij})
\hat e_{ij}\cdot  \left\{
\left( \epsilon_i {\bf v}_j + \epsilon_j {\bf v}_i \right)
+ \left( {\cal P}_{ij}{\bf v}_i + {\cal P}_{ji}{\bf v}_j \right)
\right. \nonumber \\
&& \left.
- \left( \tilde \pi_{ij}\cdot {\bf v}_i + 
\tilde \pi_{ji} \cdot {\bf v}_j \right)
+ \left( {\cal J}_{ij} + {\cal J}_{ji} \right) \right\} .
\label{dene}
\end{eqnarray}

By construction, the total mass, total momentum and total energy
are conserved. This can be seen by summing the above expressions
over $i = 1, 2, \dots ,N$. 

So far, the equations are quite general and one needs constitutive
relations for the kernels of pressure, viscous stress and heat current.
For instance, the viscous stress tensor linear in the velocity gradients
may be found by discretizing Eq.(\ref{Istress}),
\begin{eqnarray}
\pi_{ij}^{\alpha \beta} &=& - \eta_i \sum_k {\cal W}(r_{jk})
 \left[ e_{jk}^\alpha v_k^\beta + e_{jk}^\beta v_k^\alpha
- {2 \over 3} \delta^{\alpha \beta} e_{jk}^\nu v_k^\nu \right] \nonumber
\\
&& - \zeta_i \delta^{\alpha \beta} \sum_k {\cal W}(r_{jk})
e_{jk}^\nu v_k^\nu \label{dstress}
\end{eqnarray}
while the heat current may be found from Eq.(\ref{Iheat})
\begin{equation}
{\cal J}_{ij} = \kappa_i \sum_k {\cal W}(r_{jk}) \hat e_{jk}
T_k .\label{dheat}
\end{equation}
One should keep in mind that in order to close the equations one
still needs the equations of state for the pressure ${\cal P}_{ij} =
{\cal P}(m_i,T_j)$ and the internal energy per particle $u_{ij} = 
u(m_i,T_j)$.

Up to here it is simply a discretization of the equations. The
interesting addition now \cite{SPH1,SPH2,DPD} is to assume that the 
positions ${\bf r}_i$ of the cells become the positions of particles
that are allowed {\it to move}. This is certainly a bold 
assumption, since making the fluid particles move
should be done in a Lagrangian formulation of the fluid dynamics
rather than in an Eulerian one. The present are a different type of
particles, however, since as we can see from the above equations, they
not only change their momenta but they also have variable mass
and carry with them their internal energy in addition to their
kinetic one. Nevertheless, one may justify it
by  arguing that one is actually looking at every instant of time
to a state of the fluid not on a grid but rather on a ``fluidized
grid''; the motion law of ${\bf r}_i$ being used as an updating
of the state of the fluid on such a grid. In any case one can asses
the validity of such an assumption {\it a posteriori}; as we
mentioned in the Introduction, successful simulations of actual flows
with schemes like the present one have been
reported, see Refs. \cite{ESPH,liver,DPD2,EDPD,Spanish,Ernst}. Although
the choice of the law of motion is arbitrary, it seems ``natural''
to consider the rate of change of the position of the cell as the velocity
of the fluid at that point:
\begin{equation}
{d {\bf r}_i(t) \over dt} = {\bf v}_i(t) . \label{motion}
\end{equation}
We point out that this choice is not unique; see Ref. \cite{SPH2}
for other forms used in SPH simulations.

In principle the above scheme is complete and closed. However, for an
actual implementation of a simulation based on it there are further
questions to resolve, such as the boundary conditions in terms of the
particles and the discretization of time. Since there are already in 
the literature a host of procedures \cite{DPD2,LE} both for dealing with
boundary conditions between particles and solid frontiers and for the
time discretization, we shall only discuss the latter because of its
relevance in the inclusion of hydrodynamic fluctuations.

A simple algorithm to simulate the dynamics consists of a two-step
propagation in time \cite{DPD}. First, there is a ``collision" step in 
which one finds the values of $m_i$, ${\bf p}_i$ and $e_i$ 
at time $t + \Delta t$ from from knowledge of all the variables
at time $t$, using Eqs. (\ref{dmass}) - (\ref{dene}), with
\begin{equation}
{\partial A_i(t) \over \partial t} \approx {A_i(t + \Delta t) - A_i(t)
\over \Delta t} \label{deriva}.
\end{equation}
It is then followed by a
``propagation" step in which the positions ${\bf r}_i(t + \Delta t)$
are computed using
\begin{equation}
{\bf r}_i(t + \Delta t) \approx {\bf r}_i(t) + 
\Delta t \> {\bf v}_i(t + \Delta t).
\end{equation}
This combination is more accurate than if both steps were done
with Eq. (\ref{deriva}). This algorithm, however, is also useful
to include the fluctuations as part of the evolution and not as
a posteriori calculation, 
being thus helpful
in determining the stability of the flow. This is an important point 
since the purpose of the simulations is to solve the equations by an 
actual propagation in time of the flow. (An analytical solution need 
not be done in this way; for instance, if the equations are linearized 
one may solve them using an integral transform technique.) Thus, one can 
include in the equations for the momentum and the energy, Eqs. (\ref{dmom})
and (\ref{dene}), discretized versions of the random viscous tensor
$\tilde \pi_{ij}^R$, and random heat current ${\cal J}_{ij}^R$, both
symmetric in ${ij}$. 
Since their second moments must obey discretized
versions of the fluctuation-dissipation expressions (\ref{flustress}) and
(\ref{fluheat}) these tensors may be added as
\begin{eqnarray}
\pi_{ij}^{R \alpha \beta}(t) &=& \left( 2 k T_i(t) \eta_j(t) + 
2 k T_j(t) \eta_i(t) 
\right)^{1/2} \Delta X_{ij}^{\alpha \beta}(t) \nonumber \\
&& + \left(k T_i(t) \zeta_j(t) + k T_j(t) \zeta_i(t) \right)^{1/2} 
\Delta Y_{ij}^{\alpha \beta}(t) \label{rand} \\
{\cal J}_{ij}^{R\alpha}(t) &=& \left( k T_i(t) \kappa_j(t)
+ k T_j(t) \kappa_i(t) \right)^{1/2} \Delta Z_{ij}^{\alpha}(t) \label{rand2}
\end{eqnarray}
where $\Delta X_{ij}^{\alpha \beta}(t)$, $\Delta Y_{ij}^{\alpha \beta}(t)$ and
$\Delta Z_{ij}^{\alpha}(t)$, symmetric in ${ij}$, represent 
independent random increments (Wienner processes \cite{Gard}) with zero mean and 
correlations
\begin{eqnarray}
\left< \Delta X_{ij}^{\alpha \beta}(t) 
\Delta X_{lm}^{\gamma \nu}(t^\prime) \right> &=&
\left( \delta^{\alpha \gamma} \delta^{\beta \nu} + 
\delta^{\alpha \nu} \delta^{\beta \gamma}
-  {2 \over 3} \delta^{\alpha \beta} \delta^{\gamma \nu} \right)
\nonumber \\
&&\times \left( \delta_{il} \delta_{jm} + \delta_{im} \delta_{jl} \right) 
\delta_{t t^\prime} \nonumber \\
\left< \Delta Y_{ij}^{\alpha \beta}(t) 
\Delta Y_{lm}^{\gamma \nu}(t^\prime) \right>&=&
\delta^{\alpha \beta} \delta^{\gamma \nu} 
\left( \delta_{il} \delta_{jm} + \delta_{im} \delta_{jl} \right) 
\delta_{t t^\prime} \nonumber \\
\left< \Delta Z_{ij}^{\alpha}(t) 
\Delta Z_{lm}^{\beta}(t) \right>&=&
\delta^{\alpha \beta} 
\left( \delta_{il} \delta_{jm} + \delta_{im} \delta_{jl} \right) 
\delta_{t t^\prime}
\end{eqnarray}
From a practical point of view, any good commercial pseudo-random
number generator suffices for these increments. The important aspect
we want to stress is that, in contrast to the continuous version 
(\ref{flustress}) and (\ref{fluheat}), here the temperature, viscosities
and thermal conductivity that appear in Eqs.(\ref{rand}) and (\ref{rand2})
are evaluated at the current values of the full fluctuating quantities 
and not at the values of the variables at the underlying flow. Since
it is assumed that the fluctuation are always small and do not make
the flow unstable, this is a minor approximation. Moreover, as mentioned
above, if the flow do become unstable by adding the random viscous
tensor and heat current that may imply, barring numerical inaccuracies,
that either the flow is indeed unstable or that the method itself does not
faithfully describe the flow.

It should be clearly understood that the particle-like representation of
a continuous fluid flow depends on two different approximation; first,
one approximates the true differential laws by integral expressions with
a finite width $r_0$ of the weighting function; and second, the integrals
are discretized. These approximations pose constraints on the length scales
of the fluid. On the one hand, the density of point-particles must be such
that the mean particle separation $(\Delta V)^{1/3}$ is smaller than $r_0$
in order to have a good approximation of the integrals;
and on the other hand, a typical hydrodynamical length, call it $\lambda$, 
must be larger than $r_0$ itself in order to have a good representation
of the gradients in terms of the integrals (i.e. an independence of the 
parameter $r_0$, see Eq.(\ref{int3}).) That is, one should always have,
\begin{equation}
(\Delta V)^{1/3} < r_0 < \lambda . \label{compara}
\end{equation}
This way, the limit $r_0 \to 0$ implies not only the equality of
the integral and differential forms of the conservation laws, but also
the continuum limit itself.	

\subsection{SPH and DPD as special cases}

The purpose of this section is not to make a revision nor 
a comparison of SPH and DPD schemes with the present one,
but rather, to show that they may be viewed as special cases of a
more general scheme that reduces to the macroscopic conservation laws
of fluids.

As we have seen the discretized conservation equations 
(\ref{dmass})-(\ref{dene}) are very general. One still needs to
provide constitutive relations for the pressure, viscous tensor and
heat current and, as long as $ij$-symmetrized forms are provided,
the conservation laws are guaranteed. The particular expressions given
in the previous section, such as (\ref{dstress}) and (\ref{dheat}), are just 
examples. But before we present other forms used, such as those of
DPD and SPH, we want to mention some aspects of the time derivatives
used.

In the schemes used in SPH and DPD, the time derivatives of the
properties of the particles have been interpreted as already including
the convective contribution. In the discretized version this is
equivalent to identify
\begin{eqnarray}
{d m_i \over dt} &=& {\partial m_i \over \partial t} - 
\sum_j {\cal W}(r_{ij}) \hat e_{ij} \cdot {\bf v}_i m_j
 \\
{d p_i^\alpha \over dt} &=& {\partial p_i^\alpha \over \partial t} - 
\sum_j {\cal W}(r_{ij}) \hat e_{ij} \cdot {\bf v}_i p_j^\alpha
 \\
{d \epsilon_i \over dt} &=& {\partial \epsilon_i \over \partial t} - 
\sum_j {\cal W}(r_{ij}) \hat e_{ij} \cdot {\bf v}_i \epsilon_j
\end{eqnarray}
This is a subtle point: One the one hand, one could argue that the derivative
is following the motion of the fluid particle, like in a Lagrangian
formulation; however, the right-hand-side of the corresponding conservation laws
(\ref{dmass})-(\ref{dene}) should be accordingly transformed. 
Since the latter is not done
in SPH and DPD, one may still say that those formulations correspond to
not too large Reynolds numbers where one can neglect the convective
contributions. It may be interesting to include those terms explicitely
in a simulation.

With the above identification of the time derivatives, we can now see a
closer resemblance to the equations of SPH and DPD. For purpose of
exemplifying the relationship, we shall only discuss the equation
for the momentum. Using the mass conservation equation and the fact
that $p_i^\alpha = m_i v_i^\alpha$, the equation for the momentum can
be written as,
\begin{equation}
m_i{d v_i^\alpha \over d t} = \sum_j {\cal W}(r_{ij}) 
e_{ij}^\beta \left\{ 
\delta^{\alpha \beta} \left( {\cal P}_{ij} + {\cal P}_{ji} \right)
- \left( \pi_{ij}^{\alpha \beta} + \pi_{ji}^{\alpha \beta} \right) 
\right\} .\label{dvel}
\end{equation}

In Refs.\cite{SPH1,SPH2,ESPH}, SPH is formulated with
forms for the pressure such as
\begin{equation}
{\cal P}_{ij} + {\cal P}_{ji} = {P_i \over m_i^a m_j^b} +
{P_j \over m_j^a m_i^b}
\end{equation}
with $a$ and $b$ constants and with a given equation of state for 
$P_i$ in terms of $m_i$ . \cite{aclara}

The viscous stress tensor of SPH and DPD may be generally written
\begin{equation}
 \pi_{ij}^{\alpha \beta} + \pi_{ji}^{\alpha \beta} =
 A(r_{ij}) e_{ij}^\alpha (v_i^\beta - v_j^\beta) + 
 B(r_{ij}) e_{ij}^\beta (v_i^\alpha - v_j^\alpha)
 \end{equation}
with appropriate choices of $A$ and $B$ 
\cite{DPD,DPD2,EDPD,Spanish,Ernst}. It is interesting to note
that this form can give rise, in the continuum limit $r_0 \to 0$, to terms 
proportional to $\nabla^2 {\bf v}$ and $\nabla (\nabla \cdot {\bf v})$;
however, one cannot independently identify the corresponding
viscosity coefficients. This is to be contrasted with the 
expression of the tensor given by equation (\ref{dstress}) where
there is an independent identification of the viscosities.

In the DPD simulations there is an additional ingredient. Namely,
that the pressure term is taken to be stochastic. Within the present
scheme this may be interpreted as including hydrodynamic fluctuations
with white noise and with a particular temperature and viscosity as
given by equation (\ref{rand}). In this regard we differ from the
interpretation of DPD equations given by Espa\~nol et al. \cite{Spanish} and
Marsh et al. \cite{Ernst}.  In that interpretation, the equations
of DPD are taken as the ``microscopic'' dynamics of the particles of
a fluid, from which the macroscopic laws are to be extracted, much 
in the spirit of Langevin and Boltzmann equations. Within that 
interpretation they have argued that the random part should be
modified in order to account for the correct fluctuation-dissipation
relation of Langevin-like equations. According to the present theory,
since the equations of motion of the particles {\it are} only an
approximation to the macroscopic equations of the fluid flow, there is no need
to modify the DPD equations. Therefore, the random contributions of DPD 
may be seen to already refer to hydrodynamic fluctuations. Moreover, if one
wishes to find the corresponding Fokker-Planck equations to the
discretized hydrodynamic equations (\ref{dmass}) - (\ref{dene}) one
can follow the theory of van Saarloos et al. \cite{vS} of non-linear
hydrodynamic fluctuations.

\section{Final Remarks}

In this article we have presented an integral form of the conservation
laws of a macroscopic classical fluid in terms of an interpolant
for the gradient of a given function of space. This form is amenable
for a discretization of space and may be interpreted in terms of the
dynamics of ``fluid particles''. To complete this discretized 
dynamics one must provide the law of motion of the position of the
particles; one may prescribe that the field velocity equals the
rate of change of the position of the particle. Within this scheme
one can easily find out the corresponding Navier-Stokes equations of
viscous flow and the Fourier law of heat conduction. Moreover, hydrodynamic
fluctuations can also be readily taken into account.

We have argued that numerical simulations currently used, known by
Smoothed Particle Hydrodynamics (SPH) \cite{SPH1,SPH2} and 
Dissipative Particle Dynamics (DPD) \cite{DPD}, and which are based 
on a particle-like simulation of a continuum
fluid, may be seen as special cases of the general formalism here
presented. In that way, one the one hand, one guarantees that the
simulations have a correct continuum limit, and on other, there is
a clear route of how to represent, in the particle dynamics, known
effects of macroscopic fluids; for instance, with the present
theory one can see how to include thermal effects, absent in DPD 
simulations.

Finally, we want to stress the potential uses of this type of schemes.
It does seem that a complication in any simulation of fluids is the
discretization of space with its concomitant difficulties of boundary
conditions; this is the more important if one is interested in
rheological fluids, such as suspensions. That is, in order to simulate
a simple flow a discretized {\it differential} scheme (e.g. finite
differences) may appear to be better than a discretized integral
version; this is because the latter makes use of the weighting
function which in turn must resemble a delta function, and therefore,
it appears that one needs more ``particles'' than points in a grid 
\cite{liver}, {\it cf.} Eq. (\ref{compara}).
However, having paid this price, there is a host of ``tricks'' and
techniques, borrowed from standard molecular dynamics, that can
be used to simulate moving boundaries and solid objects, e.g.
Lee and Edwards shear boundaries \cite{LE}, or ``freezing'' certain
number of particles to simulate a rigid body \cite{DPD}, etc. Moreover,
the inclusion of hydrodynamic fluctuations also seem to be much easier 
within a particle-like scheme than within a field-like.

\vskip0.2in\noindent
{\bf Acknowledgment}

J.M.R. thanks useful conversations with J. Dufty and M. Ernst.
V.R.-R. thanks the hospitality of the Universitat of Barcelona during
the completion of this work. We acknowledge support from CONACYT-Mexico, 
DGAPA-UNAM and DGICYT of the Spanish Government under Grant No. 
PB96-0881.


\begin{thebibliography}{99}
\bibitem{aclara} In SPH formulations, refs. \cite{SPH1,SPH2,ESPH},
the mass of the particles are given constants and the variable that obey
the continuity equation is the
mass density. Here, the mass $m_i$ is already the discretized
mass density, see Eq. (\ref{dmass}).
\end{thebibliography}
\end{document}